\documentclass[twocolumn,superscriptaddress,final]{revtex4-1}
\usepackage{graphicx}
\usepackage{color}

\newcommand{\BS}{\mathrm{Bi}_2\mathrm{Se}_3}
\newcommand{\Ef}{E_{\mathrm{F}}}
\newcommand{\tm}{t_\mathrm{max}}%{t_0}%
\newcommand{\Ima}{I_\mathrm{max}}
\newcommand{\hw}{\hbar\omega}
\newcommand{\Dt}{\Delta t}
\newcommand{\ED}{E_{\mathrm{D}}}

\begin{document}
\preprint{v1}

\title{Band resolved imaging of photocurrent in a topological insulator}

\author{H. Soifer}\affiliation{Stanford Institute for Materials and Energy Sciences, SLAC National Accelerator Laboratory, 2575 Sand Hill Road, Menlo Park, California 94025, USA}

\author{A. Gauthier}\affiliation{Stanford Institute for Materials and Energy Sciences, SLAC National Accelerator Laboratory, 2575 Sand Hill Road, Menlo Park, California 94025, USA} \affiliation{Geballe Laboratory for Advanced Materials, Departments of Physics and Applied Physics, Stanford University, Stanford, California 94305, USA}

\author{A. F. Kemper}\affiliation{Department of Physics, North Carolina State University, Raleigh, North Carolina 27695, USA}

\author{C. R. Rotundu}\affiliation{Stanford Institute for Materials and Energy Sciences, SLAC National Accelerator Laboratory, 2575 Sand Hill Road, Menlo Park, California 94025, USA}

\author{S.-L Yang}\altaffiliation[Present address: ]{Kavli Institute at Cornell for Nanoscale Science, Laboratory of Atomic and Solid State Physics, Department of Physics, and Department of Materials Science and Engineering, Cornell University, Ithaca, NY14853, USA}\affiliation{Stanford Institute for Materials and Energy Sciences, SLAC National Accelerator Laboratory, 2575 Sand Hill Road, Menlo Park, California 94025, USA} \affiliation{Geballe Laboratory for Advanced Materials, Departments of Physics and Applied Physics, Stanford University, Stanford, California 94305, USA}

\author{H. Xiong}\affiliation{Stanford Institute for Materials and Energy Sciences, SLAC National Accelerator Laboratory, 2575 Sand Hill Road, Menlo Park, California 94025, USA} \affiliation{Geballe Laboratory for Advanced Materials, Departments of Physics and Applied Physics, Stanford University, Stanford, California 94305, USA}

\author{D. Lu} \affiliation{Stanford Synchrotron Radiation Lightsource, SLAC National Accelerator Laboratory, 2575 Sand Hill Road, Menlo Park, California 94025, USA}

\author{M. Hashimoto} \affiliation{Stanford Synchrotron Radiation Lightsource, SLAC National Accelerator Laboratory, 2575 Sand Hill Road, Menlo Park, California 94025, USA}

\author{P. S. Kirchmann}\affiliation{Stanford Institute for Materials and Energy Sciences, SLAC National Accelerator Laboratory, 2575 Sand Hill Road, Menlo Park, California 94025, USA}

\author{J. A. Sobota}\email{sobota@stanford.edu}\affiliation{Stanford Institute for Materials and Energy Sciences, SLAC National Accelerator Laboratory, 2575 Sand Hill Road, Menlo Park, California 94025, USA} \affiliation{Geballe Laboratory for Advanced Materials, Departments of Physics and Applied Physics, Stanford University, Stanford, California 94305, USA} \affiliation{Advanced Light Source, Lawrence Berkeley National Laboratory, Berkeley, California 94720, USA}

\author{Z.-X. Shen}\email{zxshen@stanford.edu}\affiliation{Stanford Institute for Materials and Energy Sciences, SLAC National Accelerator Laboratory, 2575 Sand Hill Road, Menlo Park, California 94025, USA} \affiliation{Geballe Laboratory for Advanced Materials, Departments of Physics and Applied Physics, Stanford University, Stanford, California 94305, USA}

\date{\today}

\begin{abstract}
We study the microscopic origins of photocurrent generation in the topological insulator $\BS$ via time- and angle-resolved photoemission spectroscopy. We image the unoccupied band structure as it evolves following a circularly polarized optical excitation and observe an asymmetric electron population in momentum space, which is the spectroscopic signature of a photocurrent. By analyzing the rise times of the population we identify which occupied and unoccupied electronic states are coupled by the optical excitation. We conclude that photocurrents can only be excited via resonant optical transitions coupling to spin-orbital textured states. Our work provides a microscopic understanding of how to control photocurrents in systems with spin-orbit coupling and broken inversion symmetry.
\end{abstract}

\maketitle

Topological insulators are bulk insulators with robust surface conductivity mediated by topological surface states (TSS) \cite{Fu2007,Zhang2009,Xia2009,Chen2009}. Much of the excitement surrounding topological insulators stems from the rich phenomena enabled by the spin-orbital texture of the TSS, in which the electron angular momentum is locked perpendicular to its crystal momentum $k$ \cite{Hsieh2009a, Zhang2013a,Zhu2013}. However, accessing the intrinsic properties of the TSS via conventional transport methods has been hindered by the high residual conductivity of the bulk \cite{Butch2010,Skinner2012,Barreto2014}. In contrast, light-based probes have proven fruitful because optical selection rules permit direct coupling to the spin-orbital degrees of freedom, and are therefore less obstructed by the bulk response \cite{Jozwiak2013,Zhu2014,Sanchez-Barriga2014,Sanchez-Barriga2016}. The circular photogalvanic effect bridges optics and transport phenomena: circularly-polarized light generates photocurrents which can be measured in conventional transport configurations \cite{Hosur2011,Junck2013,McIver2011,Olbrich2014,Pan2017}. The correlation of the photocurrent direction with the light helicity has been taken as evidence that the current is carried by the TSS \cite{McIver2011, Panna2017,Kastl2015,Wang2016,Braun2016}

Microscopically, the total photocurrent density $j$ parallel to the surface can have contributions from all bands $n$, in the Brillouin zone $(-k_\mathrm{BZ},k_\mathrm{BZ})$:

\begin{equation}
j = \sum_{n} j_n = -\frac{e}{\hbar} \sum_{n} \int_{0}^{k_\mathrm{BZ}} \left[ \rho_n(+k) - \rho_n(-k)\right] \frac{\partial \varepsilon_n}{\partial k}dk ,
\label{eq:current}
\end{equation}
which shows that a momentum asymmetry in the electron density $\rho_n(k)\neq\rho_n(-k)$ is required for generating current \cite{Gudde2007}.
Individual band contributions $j_n$ can be directly measured using time- and angle- resolved photoemission spectroscopy (trARPES) which resolves the photoexcited populations along the electronic band dispersions $\varepsilon_n(k)$. Early trARPES works have demonstrated that circularly-polarized light generates an asymmetric electron distribution in the TSS \cite{Sanchez-Barriga2016, Kuroda2016, Kuroda2017, Bugini2017}. The question remains whether this phenomenon is exclusive to topological states or can be generalized to other spin-orbit coupled states. It is therefore important to characterize the photoexcited carriers over a broader energy range, including photoholes injected below the Fermi level \cite{Pan2017}. A detailed understanding of optical coupling to spin-orbital degrees of freedom is required for a microscopic picture of photocurrent generation in topological insulators.

\begin{figure*}
	\includegraphics[width = \textwidth]{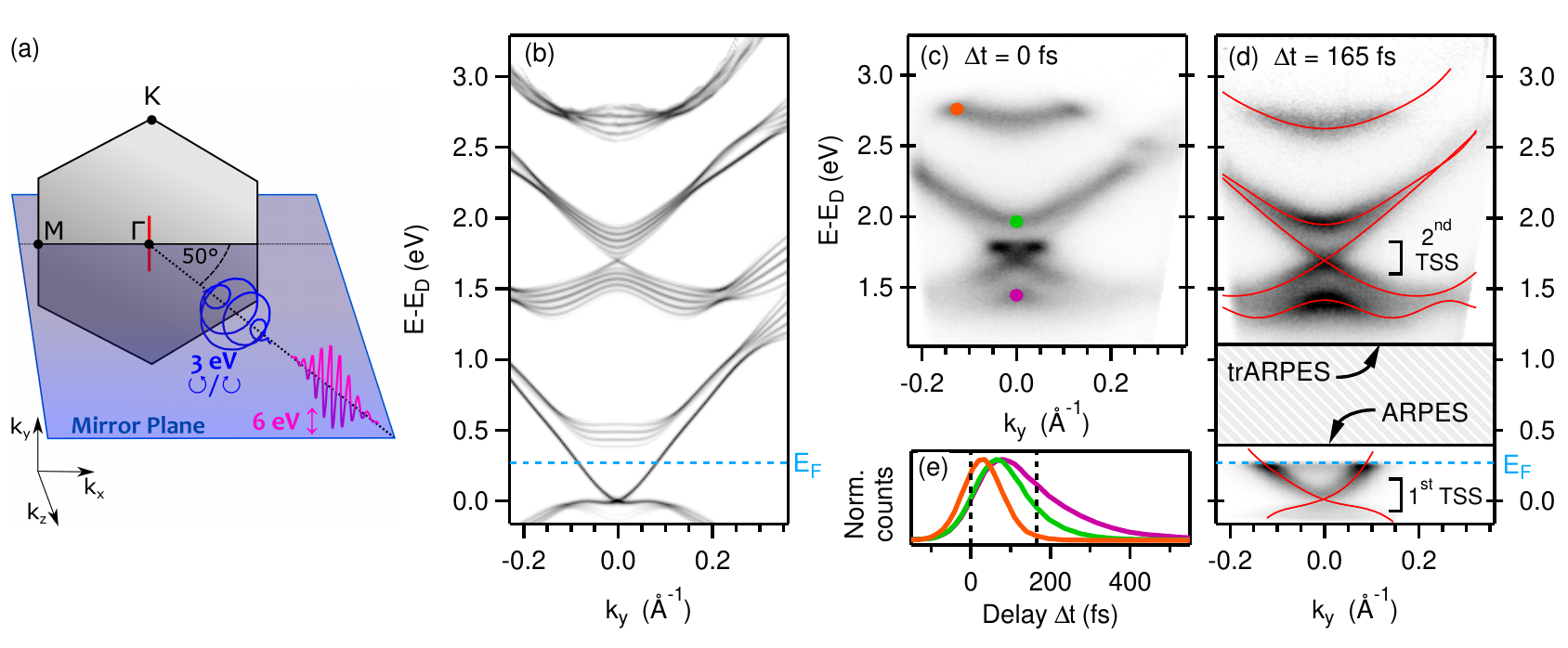}%
	\caption{\textbf{Unoccupied band structure of $\mathbf{Bi_2Se_3}$.} (a) Sketch of the experimental geometry. The cut direction is shown in red. (b) DFT calculation of the band structure. Energies are referenced to the $1^\mathrm{st}$ Dirac point. The blue dashed line marks the Fermi level ($\Ef-\ED=270$\,meV). (c) and (d) Population of unoccupied bands from trARPES at two delays (average of population excited by $\sigma_+$ and $\sigma_-$).  The bottom panel of (d) shows the the first Dirac cone, measured with 6\,eV ARPES. The intensities in the top panel of (d) were rescaled exponentially as a function of energy for better visibility. Red lines are guides to the eye that trace the bands. (e) Transient intensity for 3 spectral regions, as marked by dots of the same color in (c). Dashed vertical lines mark the delays at which the trARPES spectra in (c) and (d) were taken.\label{Fig1}}
\end{figure*}

In this Letter, we use trARPES to study the momentum-asymmetric optical transitions responsible for photocurrent generation in $\BS$. We excite a wide range of states, and analyze the population dynamics to simultaneously yield the momentum distributions of electrons and holes. Our results vividly highlight the key role of resonant transitions in generating asymmetric distributions using circularly-polarized light. Moreover, we find that population asymmetries are not exclusive to topological states, and broadly reexamine the ingredients required for generating such asymmetries.

In our trARPES experiments we used circularly-polarized 3.02~eV pump photons to populate unoccupied bands above the Fermi level, $\Ef$, and s-polarized 6\,eV  probe pulses to photoemit electrons at a variable delay $\Dt$. The overall time resolution of 100\,fs is extracted from cross-correlations of pump and probe pulses. $\Dt=0$ refers to both pulses overlapping in time. The incidence plane of light is along the mirror plane of the sample, see Fig.~\ref{Fig1}(a). Photoelectrons are collected in a hemispherical analyzer along the $\Gamma$-K direction \cite{Sobota2012}. Additional ARPES data of the equilibrium band structure was recorded at beamline 5-2 of the Stanford Synchrotron Radiation Lightsource using 35\,eV photons. We perform density functional theory (DFT) calculations using the VASP \cite{kresse1993ab, kresse1994ab, kresse1996efficiency, kresse1996efficient, kresse1994norm, kresse1999ultrasoft} package using PAW PBE \cite{blochl1994projector, kresse1999ultrasoft, perdew1996generalized, perdew1996erratum} GGA-type pseudopotentials. The calculation used a 7 quintuple layer slab geometry based on the experimental coordinates and with a 30\,{\AA} vacuum layer.  We used a $15\times15\times1$ momentum grid centered on $\Gamma$, and a wave function cutoff of 250\,eV.  Spin-orbit coupling was included through a non-self-consistent calculation after a converged density was obtained. The energy scale was renormalized by a factor of 1.11 to match the experimental energy difference between the $1^\mathrm{st}$ and $2^\mathrm{nd}$ Dirac points of 1.7\,eV.

Figure~\ref{Fig1}(b) shows the calculated band structure of $\BS$ with the cone-shaped dispersion of the $1^\textrm{st}$ TSS visible at the bottom \cite{Zhang2009}. Throughout the paper we reference energies to the 1$^\mathrm{st}$ Dirac point, $E_\mathrm{D}$. The blue dashed line marks $\Ef$, which is 270\,meV above $\ED$ in our experiments. A second cone-shaped dispersion at 1.7\,eV corresponds to the $2^\textrm{nd}$ TSS above $\Ef$ \cite{Niesner2012, Sobota2013}. 

\begin{figure}
	\includegraphics[width=\columnwidth]{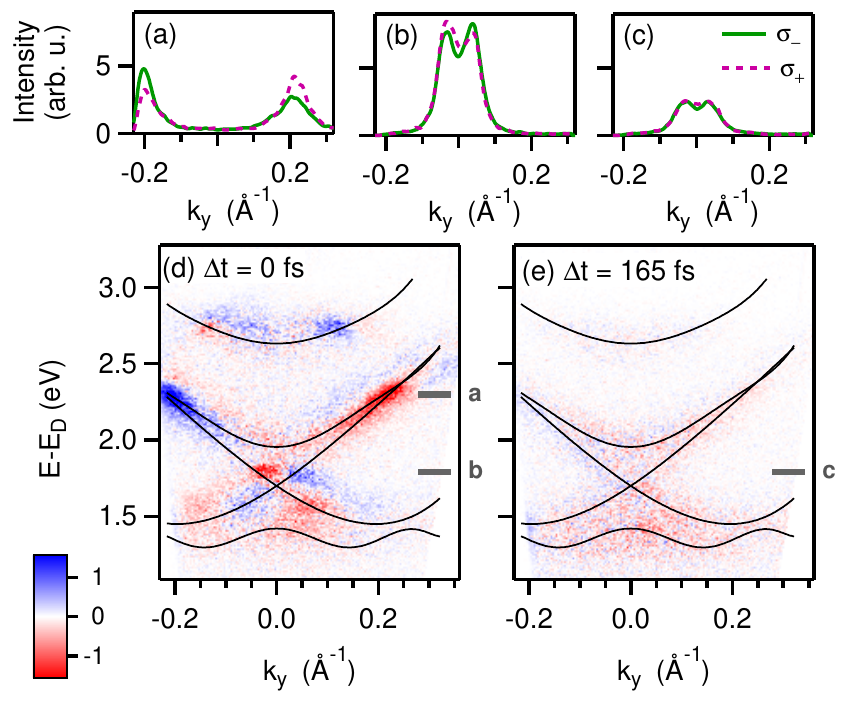}%
	\caption{\textbf{Helicity Dependent Population Asymmetry.} (a)-(c) Momentum Distribution Curves (MDCs) of the unoccupied band structure, excited with $\sigma_+$ (dashed-purple) and $\sigma_-$ (solid-green) polarized pulses. (a) and (b) were taken at $\Dt=0$, and (c) at $\Dt=165$\,fs, at the energies marked by the short gray lines in (d)-(e). (d),(e) Asymmetry image: difference between the populations of the unoccupied bands when excited by $\sigma_-$ and $\sigma_+$ polarized pulses, taken at $\Dt=0$ and 165\,fs respectively. Black lines are guides to the eye that follow the dispersions of the unoccupied bands.\label{Fig_CD}}
\end{figure}

\begin{figure*}
	\includegraphics[width = \textwidth]{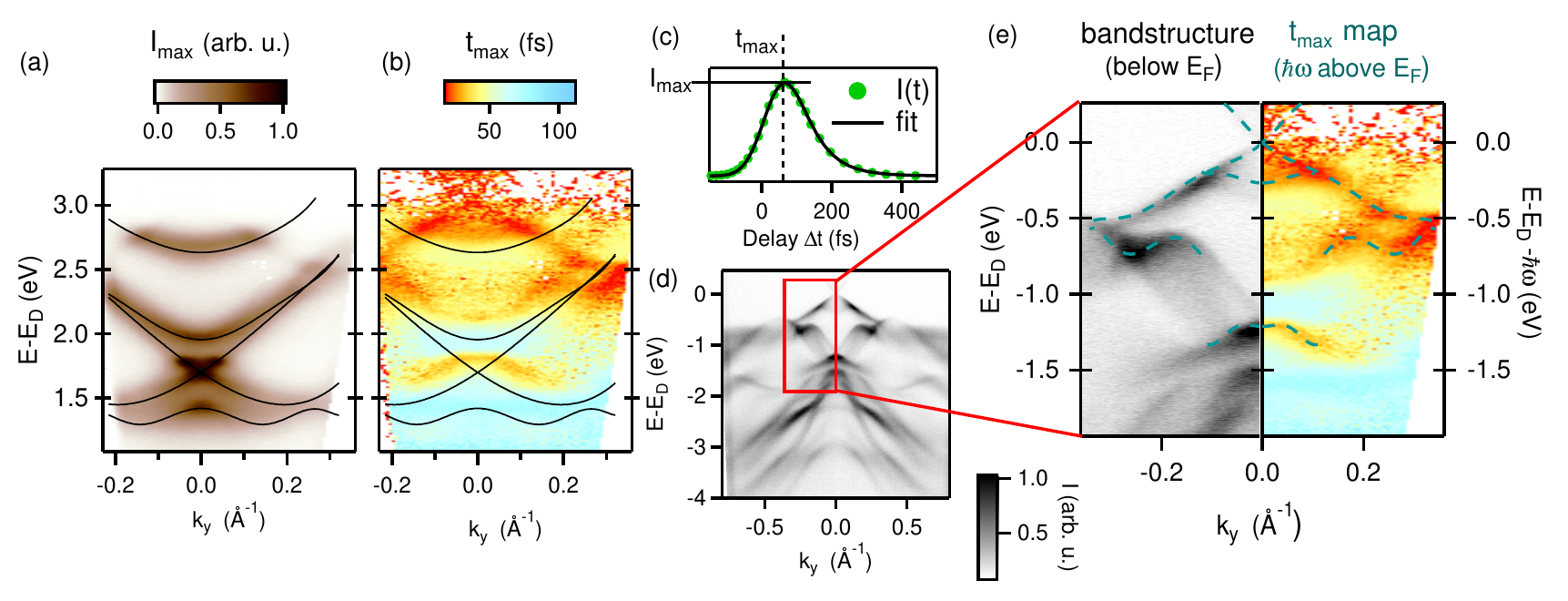}%
	\caption{\textbf{Time Mapping: $E$ and $k$ resolved analysis of the population dynamics.} (a) $\Ima(k,E)$: Maximum intensity from fits of each energy-momentum bin. (b) $\tm(k,E)$: delay at which the intensity is maximal for each bin. Solid lines in (a) and (b) represent the dispersions of the unoccupied bands. Bins where the fit did not converge appear white in both images. (c) Prototypical transient population (green dots, same as green curve in Fig.~\ref{Fig1}(e) ),  and fit (solid black). $\Ima$ and $\tm$ are shown. (d)  Occupied band structure from ARPES, using s-polarized 35\,eV photons. (e) Comparison between $\tm$ map (right) and the occupied states one pump photon energy below ($\hw=3.02$\,eV) (left). The dashed lines are guides to the eye based on the early-rise times in the $\tm$ map.\label{Fig2}}
\end{figure*}

The population in the unoccupied bands measured at $\Dt=0$ and 165\,fs is shown in Fig.~\ref{Fig1}(c) and (d) respectively. The spectrum at $\Dt=165$\,fs clearly corresponds to the calculated bands. Red lines are guides to the eye that trace the band dispersions and are used throughout the paper as reference to the unoccupied bands. The first TSS is shown at the bottom of Fig.~\ref{Fig1}(d), measured via 6\,eV ARPES.  
The spectral intensity distribution at  $\Delta t = 0$ is different, in particular the increased intensity around the $2^\mathrm{nd}$ TSS at $E-E_D=1.8$\,eV. The intensity distribution at $\Dt=0$ is determined by the joint density of states between the initial and final states of the electronic excitation \cite{Sobota2013,Yang2017}, whereas the distribution at later delays, after the electrons have scattered, reflects more directly the unoccupied band structure. The normalized, transient intensity in three different spectral regions is shown in Fig.~\ref{Fig1}(e).

Next, we explore how the momentum distribution of the photoexcited electrons depends on the light helicity. Figure~\ref{Fig_CD}(a-c) displays momentum distribution curves (MDCs) of the unoccupied states at two energies and two delays, when excited by $\sigma_+$ (dashed purple) and $\sigma_-$ (solid green) circular polarizations. At $\Dt=0$ and $E-\ED=2.3$\,eV [Fig.~\ref{Fig_CD}(a)] circular polarization excites an asymmetric population distribution, indicating that at this energy there is a net flow of electrons in one direction. This asymmetry reverses direction when switching the light helicity. Similarly, in Fig.~\ref{Fig_CD}(b) ($\Dt=0$, $E-\ED=1.8$\,eV) we observe the signature of a helicity dependent asymmetry, whereas the MDCs at the same energy and $\Dt=165$\,fs display symmetric populations [Fig.~\ref{Fig_CD}(c)].

The difference plots between the $\sigma_+$ and $\sigma_-$ excited spectra are shown in Fig.~\ref{Fig_CD}(d)-(e), where the red/blue colors signify a helicity-dependent population asymmetry. The difference map at $\Dt=0$ in Fig.~\ref{Fig_CD}(d) reveals a complex pattern of band- and momentum-dependent asymmetries which decay within 165\,fs [Fig.~\ref{Fig_CD}(e)]. These results show that multiple bands contribute to the photocurrent, and can contribute with opposite directions. 

\begin{figure}
	\includegraphics[width = \columnwidth]{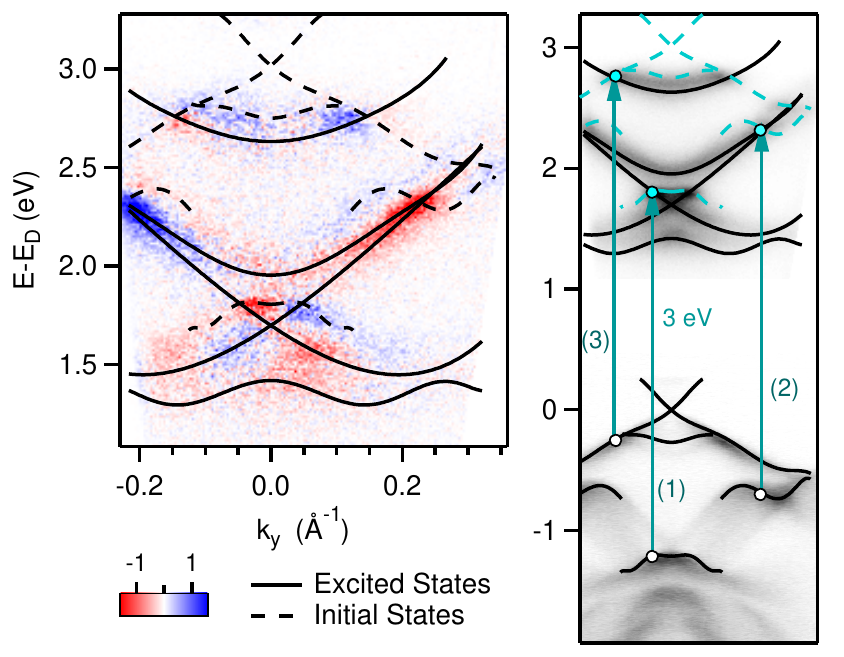}%
	\caption{\textbf{Resonant transitions.} (a) Asymmetry image: difference between the populations of the unoccupied bands when excited by $\sigma_-$ and $\sigma_+$ polarized pump, at $\Dt=0$\,fs. [same as Fig.~\ref{Fig_CD}(d)]. The solid lines  follow the dispersions of the unoccupied bands. Dashed lines follow the early-rise times from the $\tm$ map, marking the dispersions of the initial states $\hw=3.02$\,eV below. (b) Schematic visualization of the resonant transitions between occupied and unoccupied states of $\BS$ with a 3\,eV excitation. The solid black lines mark the bands, and the blue-dashed lines mark the initial states upshifted by the excitation.\label{Fig3}}
\end{figure}

It is important to note that the photo-excited electrons do not represent all of the charge carriers contributing to the current, the remainder being the corresponding holes injected below $\Ef$.
Though we do not have direct access to the hole population, the conservation of energy and momentum in optical transitions links the electron and hole distributions. Therefore, we can infer the hole distribution by determining which part of the excited population corresponds to direct optical transitions, and which to filling via secondary scattering processes. These two processes can be distinguished by their timescales, with direct optical transitions causing a faster population rise than secondary scattering \cite{Weinelt2004,Reimann2014}.

We analyze the rise times of the photoexcited population to identify which parts of the unoccupied bands are populated via direct optical transitions. This time-mapping analysis is based on energy- and momentum-resolved fits of the intensity dynamics: for each spectral bin (0.007\,\AA$^{-1}$ x 12\,meV), we fit the transient signal with a Gaussian convoluted with an exponential decay to model the filling and subsequent decay of the optically excited population. A typical fit is shown in Fig.~\ref{Fig2}(c). We focus on two fitted quantities: the maximum intensity, $\Ima$, and the delay at which the maximum intensity is reached, $\tm$, which are plotted in Fig.~\ref{Fig2}(a) and (b) respectively. The intensity $\Ima$ bears a close similarity to the spectral intensity at $\Delta t = 0$ [Fig.~\ref{Fig1}(b)], which however is not the case for the $\tm$ map. It can be separated into two types of regions: where the intensity rises early ($\tm \lesssim 50$\,fs, yellow-red regions), and regions with later rise times (blue-white). This observation underlies our approach to distinguish between direct optical transitions and secondary scattering using population rise times.

Though the early-rise regions do not follow the dispersion of the unoccupied bands (solid black lines), we find that they reflect the initial states of the excitation (occupied states below $\Ef$). To demonstrate this link, we measure the occupied band structure of $\BS$ via ARPES as shown in Fig.~\ref{Fig2}(d). 
The $\tm$ map agrees remarkably well with the occupied bands, when shifted down by one pump photon energy ($\hbar\omega=3.02$\,eV), see Fig.~\ref{Fig2}(e). 
Guides to the eye (dashed blue) trace the early-rise features in the $\tm$ map and are reflected around $k=0$ for an easier comparison. Notably, for every early-rise feature in the time-map there is a corresponding band in the initial state spectrum, while the opposite is not true - there are occupied bands without a corresponding early-rise feature in the $\tm$ map (see e.g., electronic band at $E-\ED=-1.5$\,eV). We conclude that the rise-time mapping reveals only the occupied states which participate in the optical excitation process.

This mapping of initial states allows us to characterize the transitions which result in asymmetric populations. Figure~\ref{Fig3}(a) shows the asymmetry image at $\Dt=0$, overlaid with the dispersions of the excited states (solid lines), as well as the initial states obtained from our time-mapping analysis (shifted up by $\hw$, dashed lines). These three elements together reveal that the regions characterized by the strongest electron population asymmetry are those where the initial and excited states overlap in momentum and energy. Only resonant optical transitions lead to asymmetric populations and consequently contribute to photocurrent generation. Scattering processes subsequently fill non-resonant states but lead to a loss of asymmetry, in agreement with the symmetric populations observed at later delays [Fig.~\ref{Fig_CD}(e)].

We can use this information to describe the contribution of the photoholes to the current. Since the measured asymmetric population is associated with energy- and momentum-conserving direct optical transitions, it also reflects the asymmetry of the photohole distribution, shifted by the pump photon energy. Despite having similar populations, the magnitudes of the electron and hole contributions to the photocurrent are not equal, since they originate in bands with different velocities $\frac{\partial \varepsilon_n}{\partial k}$ (see Eq. ~\ref{eq:current}). The power of our approach is that we directly extract the band velocities of the electrons from the trARPES spectrum at later delays and of the holes from the dispersions via time-mapping. For example, the contribution to the photocurrent from the transition into the second TSS [arrow (1) in Fig.~\ref{Fig3}(b), for $\sigma_+$ excitation] is due to electrons populating the left, negatively dispersing, branch of the unoccupied Dirac cone, whereas holes reside in a weakly dispersing region of the bulk band structure and contribute negligibly to the photocurrent. We therefore expect that the photocurrent contribution from this particular transition is dominated by electrons in the unoccupied TSS flowing in the $-y$ direction.

Not each asymmetric distribution is associated with a TSS, as evidenced by transition (2). This observation motivates us to examine the fundamental mechanism for driving $k$-asymmetric optical excitations. Circularly polarized light couples to orbital degrees of freedom by incrementing or decrementing the orbital angular momentum $m_L$ along the quantization axis defined by the light propagation vector. Though spin-orbit coupling entangles $m_L$ with the electron momentum $k$ \cite{Zhang2013a,Zhu2013}, this alone is insufficient for enabling helicity-dependent transitions because $\pm m_L$ are degenerate in the presence of inversion symmetry.  This degeneracy is lifted at the surface \cite{Park2011,Sunko2017}, while time-reversal symmetry ensures that states at $\pm k$ have opposite signs of $m_L$. Thus, the key ingredients for enabling $k$-asymmetric optical excitations are spin-orbit coupling and inversion-symmetry breaking. Though transitions (1) and (3) involve a TSS, it is not obvious that transition (2) involves states of broken inversion symmetry. We hypothesize that the final state includes a topologically trivial surface resonance with a spin-orbital texture oriented opposite to that of the TSS, as was recently observed for states accompanying the first TSS \cite{Cacho2015,Jozwiak2016}.

Since both topological and trivial states participate in asymmetric optical excitations, this phenomenon should be abundant in many families of spin-orbit coupled materials with broken inversion symmetry. Indeed, the same concepts underlie valley-dependent optical transitions in transition metal dichalcogenides \cite{Mak2012}, as well as proposals to manipulate the chirality of Weyl fermions in semimetals \cite{Yu2016,Ma2017}. In fact, the photocurrent associated with these transitions in Weyl semimetals is of central importance, since it is currently the only quantized observable predicted for this class of materials \cite{DeJuan2017}.

This work establishes trARPES as a powerful toolset for identifying the optical transitions which form the microscopic basis for photocurrent generation via the circular photogalvanic effect. We account for contributions from photoelectrons and photoholes, including both topological and trivial states. An important finding is that different bands contribute to the net current with opposing signs. Therefore, there are promising opportunities to optimize the magnitude of the photocurrent for applications by exploring resonance conditions achieved by other photon energies. We also point out that we excite photocurrents in the 2$^\mathrm{nd}$ TSS using easily generated light in the visible range. These results lay the foundation for future measurements of the spin-polarization of the excited carriers, which would be a milestone toward establishing complete control over the degrees of freedom which make topological materials attractive for spintronic applications. The concepts exemplified here apply universally to materials with spin-orbit coupling and inversion-symmetry breaking, and are therefore likely to find relevance for optical investigations of transition-metal dichalcogenides and Weyl semimetals.

\subsection*{Acknowledgments}
This work was supported by the U.S. Department of Energy, Office of Science, Basic Energy Sciences, Materials Sciences and Engineering Division. Stanford Synchrotron Radiation Light source is operated by the U.S. Department of Energy, Office of Science, Office of Basic Energy Sciences. A.G. acknowledges support from the National Defense Science and Engineering Graduate Fellowship program. H.X. and J.A.S. were in part supported by the Gordon and Betty Moore Foundation's EPiQS Initiative through Grant No. GBMF4546. S.-L.Y. acknowledges support by the Stanford Graduate Fellowship, and the Kavli Postdoctoral Fellowship at Cornell University.

\bibliography{BiSe_refs2}

\end{document}